# Orbital Ordering in the Charge Density Wave Phases of $BaNi_2(As_{1\text{-}x}P_x)_2$

Tom Lacmann,[1,2,*] Robert Eder,[1] Igor Vinograd,[1,†] Michael Merz,[1,3] Mehdi Frachet,[1,‡] Philippa Helen McGuinness,[1] Kurt Kummer,[4] Enrico Schierle,[5] Amir-Abbas Haghighirad,[1] Sofia-Michaela Souliou,[1] and Matthieu Le Tacon[1,§]

[1] *Institute for Quantum Materials and Technologies, Karlsruhe Institute of Technology, Kaiserstraße 12, 76131 Karlsruhe, Germany*
[2] *Laboratory for Quantum Magnetism, Institute of Physics, École Polytechnique Fédérale de Lausanne (EPFL), 1015 Lausanne, Switzerland*
[3] *Karlsruhe Nano Micro Facility (KNMFi), Karlsruhe Institute of Technology, Kaiserstraße 12, 76131 Karlsruhe, Germany*
[4] *ESRF, The European Synchrotron, 71, avenue des Martyrs, CS 40220 F-38043 Grenoble Cedex 9, France*
[5] *Helmholtz-Zentrum Berlin für Materialien und Energie, Berlin 12489, Germany*

We use resonant x-ray scattering at the nickel $L_{2,3}$ edges to investigate the interplay between orbital degrees of freedom and charge density waves (CDWs) in the superconductor $BaNi_2(As_{1\text{-}x}P_x)_2$. Both the incommensurate and commensurate CDWs in this system exhibit strong resonant enhancement with distinct energy and polarization dependencies, indicative of orbital ordering. Azimuthal-angle-dependent measurements reveal a lowering of the local Ni site symmetry, consistent with monoclinic or lower point group symmetry. The scattering signatures of both CDWs are dominated by contributions from Ni $d_{xz,yz}$ orbitals, with similar orbital character despite their distinct wave vectors. These findings point to a shared orbital-driven formation mechanism and provide new insight into the symmetry breaking and orbital and nematic fluctuations in the high-temperature regime of the superconductor $BaNi_2(As_{1\text{-}x}P_x)_2$.

*Introduction*—Orbital degrees of freedom are a fundamental aspect of the physics of quantum materials, influencing a wide range of phenomena from unconventional superconductivity to charge and spin order. While their role is well established in strongly correlated oxides [1,2], recent studies have highlighted their relevance in metallic systems, where orbital-selective effects can lead to enhanced correlations, electronic nematicity, and exotic ordered states [3,4]. Transition-metal pnictides provide a compelling platform to explore these effects due to their multiorbital nature and moderate electronic correlations [5].

$BaNi_2As_2$, a structural relative of the parent compound of Fe-based superconductors $BaFe_2As_2$ [6,7], has recently emerged as an interesting material in this context. It exhibits a complex phase diagram involving charge-density-wave (CDW) orders, nematicity and superconductivity ($T_c = 0.7$ K) that can be tuned using chemical substitution [8–13] or pressure [14]. Above the onset of long-range order, strong fluctuations of an incommensurate CDW (I-CDW) have been reported in the high-temperature tetragonal phase $BaNi_2As_2$ [15,16], accompanied by an anomalous splitting of the doubly degenerate $E_g$ phonon mode [17]. This behavior has been interpreted as evidence of strong coupling of the lattice vibrations to slow nematic fluctuations, likely driven by orbital degrees of freedom in the absence of magnetism. The unconventional nature of the electron-lattice coupling in $BaNi_2As_2$ is further emphasized by the fact that the wave vector $q_{\mathrm{I-CDW}}$, at which the soft phonon mode (dispersing from the anomalous $E_g$ mode at $q = 0$) condensates, does not correspond to any Fermi surface nesting vector, nor seems to be dictated by the anisotropy of the calculated electron-phonon coupling [15]. Moreover, the full softening substantially precedes the transition to the long-range I-CDW order and the subtle (fourfold symmetry breaking) orthorhombic distortion that accompanies it [16,18]. As temperature decreases, a commensurate CDW (C-CDW) emerges, concomitant with a pronounced redistribution of the population from the $d_{xy}$ to the $d_{xz,yz}$ nickel orbitals [18], underscoring the role of orbital physics in the formation of charge order.

Resonant elastic and inelastic x-ray scattering (REXS and RIXS) have proven to be powerful techniques for investigating the structure and collective dynamics of both orbital orders and CDWs in the cuprates [19,20], nickelates [21,22], and other correlated materials [23–26]. As an element- and orbital-sensitive probe, REXS directly detects CDW scattering and provides insight into its electronic and structural components [19–21]. The resonance profile of the CDW differs depending on whether the distortion is purely lattice-driven or if there is orbital order [27–29], which includes the spatial modulation of the orbital occupation or the energy of the orbital states. Azimuthal dependence measurements offer further symmetry-resolved characterization of charge and spin correlations [30,31].

*Contact author: Tom.Lacmann@epfl.ch
†Present address: Université Grenoble Alpes, INSA Toulouse, Université de Toulouse Paul Sabatier, CNRS, LNCMI, F-38000 Grenoble, France.
‡Present address: CNRS, Université Grenoble Alpes, Institut Néel, 38042 Grenoble, France.
§Contact author: Matthieu.LeTacon@kit.edu

In this Letter, we investigate whether both the I-CDW and C-CDW in pristine and phosphorus substituted $BaNi_2As_2$ exhibit orbital order and identify the orbitals involved in the CDW formation. Figure 1(a) shows a simplified phase diagram indicating the investigated substitutions. To this end, we perform energy-, polarization- and azimuthal-dependent REXS measurements at the nickel $L_{2,3}$ edge. Our results reveal a strong resonant enhancement of both the I-CDW and C-CDW signals, accompanied by pronounced polarization and azimuthal dependencies. These findings provide direct evidence for orbital order, with Ni $d_{xz,yz}$ orbitals playing a dominant role in the CDW formation. Furthermore, the observed azimuthal dependence reveals a lowering of the local Ni symmetry to at least monoclinic in the I-CDW phase. Remarkably, despite their distinct wave vectors, the I-CDW and C-CDW share a similar orbital character, pointing to a common underlying formation mechanism and providing strong evidence for the commensurate lock-in nature of the C-CDW transition. More broadly, identifying orbital-driven symmetry breaking in $BaNi_2(As_{1\text{-}x}P_x)_2$ provides a reference for understanding how orbital polarization couples to lattice and charge degrees of freedom in Ni-based pnictides and other multiorbital systems, where intertwined orbital, lattice, and electronic orders govern transport and nematic behavior.

*Experimental configuration*—REXS experiments were carried out at the UE46-PGM-1 beamline at BESSY II and complemented by REXS and high-resolution RIXS measurements at the ID32 beamline of the ESRF. Details of the experimental methods, crystal growth, and characterization of the investigated samples are provided in the Supplemental Material (SM) [32].

In the scattering experiments, both the I-CDW and C-CDW phases are characterized by sharp and intense superstructure reflections. To access these reflections and perform an azimuthal rotation around the corresponding scattering wave vector in a REXS geometry, the scattering configurations must be carefully optimized, as illustrated in Figs. 1(c), 1(d), and S1 [32]. We observed a strong scattering signal at (0 0.28 0) (I-CDW) and $(0\,1/3\,\overline{1/3})$ that requires measurements on the side of the platelike samples. Our measurements reveal a temperature dependence similar to that previously observed out of resonance with hard x-rays [Fig. 1(b)]. No scattering signal can be observed for the second I-CDW position—(0 0.28 1)—which can be accessed in resonance (see SM [32]).

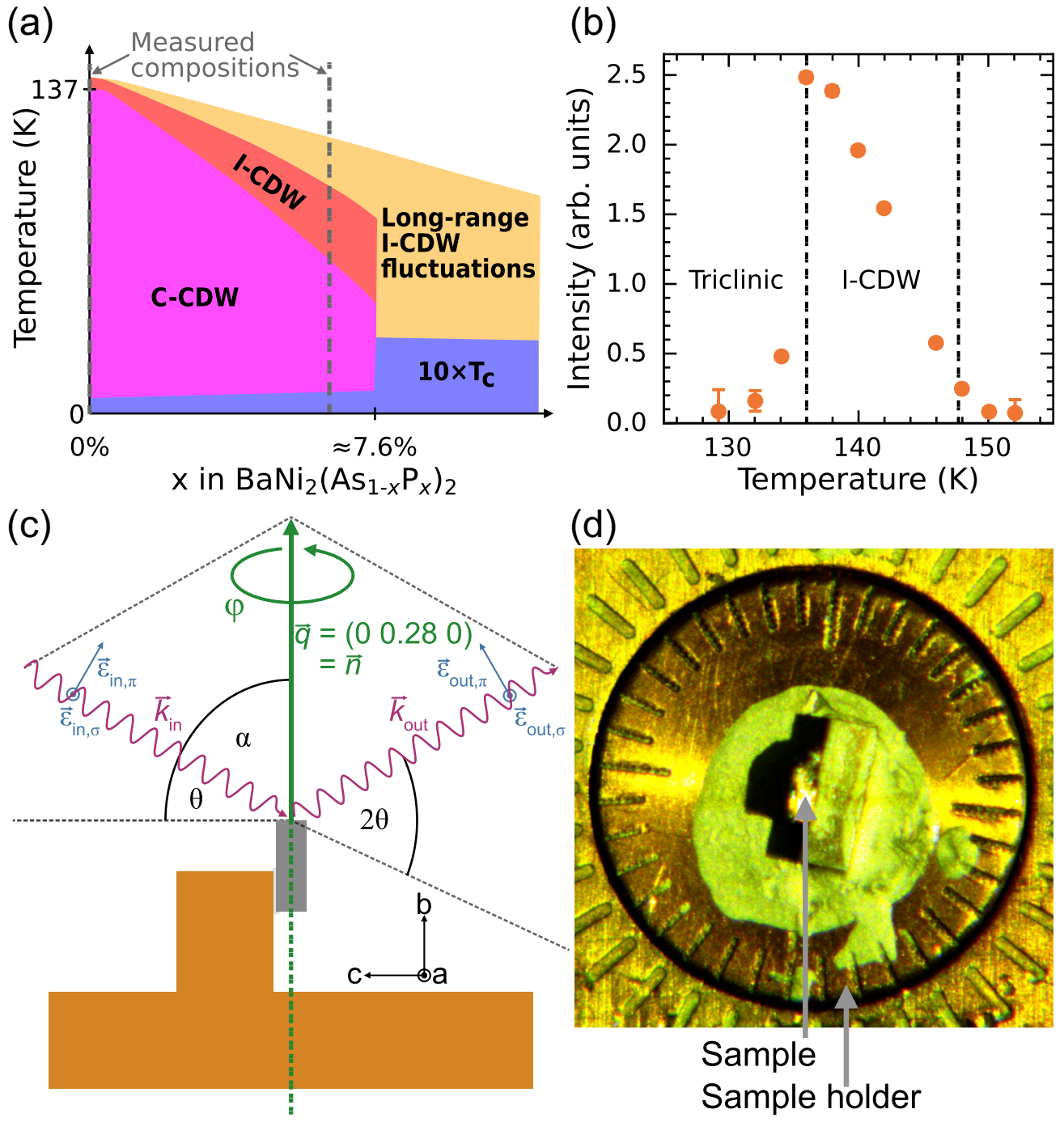

FIG. 1. (a) Simplified phase diagram of $BaNi_2(As_{1\text{-}x}P_x)_2$ indicating the I-CDW and triclinic or C-CDW phase. The two investigated substitutions are indicated by dashed gray lines. (b) Temperature dependence of the integrated intensity of the I-CDW measured in resonance. The dashed lines indicate the onset of long-range order, as determined by Lacmann *et al.* [13], and the triclinic transition. (c) Sketch of the scattering geometry used for the azimuthal-dependent REXS measurements of the I-CDW. The scattering vector (green), the incoming and scattered x-rays (purple), and the relevant angles are indicated. The azimuthal rotation (around the scattering vector), which corresponds to the angle $\varphi$ in an Eulerian cradle, is also indicated. (d) Image of Sample 1, as measured at BESSY II, mounted on the sample holder without the additional wedge.

*Resonance profiles*—We first discuss the incident photon energy dependence of the I-CDW and C-CDW scattering signals measured at the Ni $L_3$ and $L_2$ edges in a pristine $BaNi_2As_2$ single crystal. The corresponding resonances are presented in Fig. 2(a) alongside the fluorescence signal, which primarily reflects the x-ray absorption. The extracted absorption profile is consistent with that reported in a previous study [18]. Additional energy-dependent measurements for the other samples are provided in SM [32].

The resonance profiles were measured at 140 and 10.5 K corresponding to the static I-CDW and C-CDW phases, respectively. The fluorescence signal was measured at 138.5 K (on a different sample, from the same batch) just above the triclinic transition. Both the I-CDW and C-CDW exhibit a clear and nearly identical resonance profile, with a pronounced peak around 852.8 eV, located on the rising edge of the x-ray absorption spectrum. At higher energies, around 854 eV, a weak shoulder is visible in the CDW scattering signals. This shoulder shows a slight polarization dependence. Comparing the peak positions with the temperature-dependent absorption changes reported in [18], the main resonance falls within the energy range associated with the $d_{xz,yz}$ orbitals (highlighted in light red). In contrast, the shoulder appears in the energy region attributed to the Ni $d_{xy}$ orbitals (highlighted in light blue) and is significantly weaker than the main resonance.

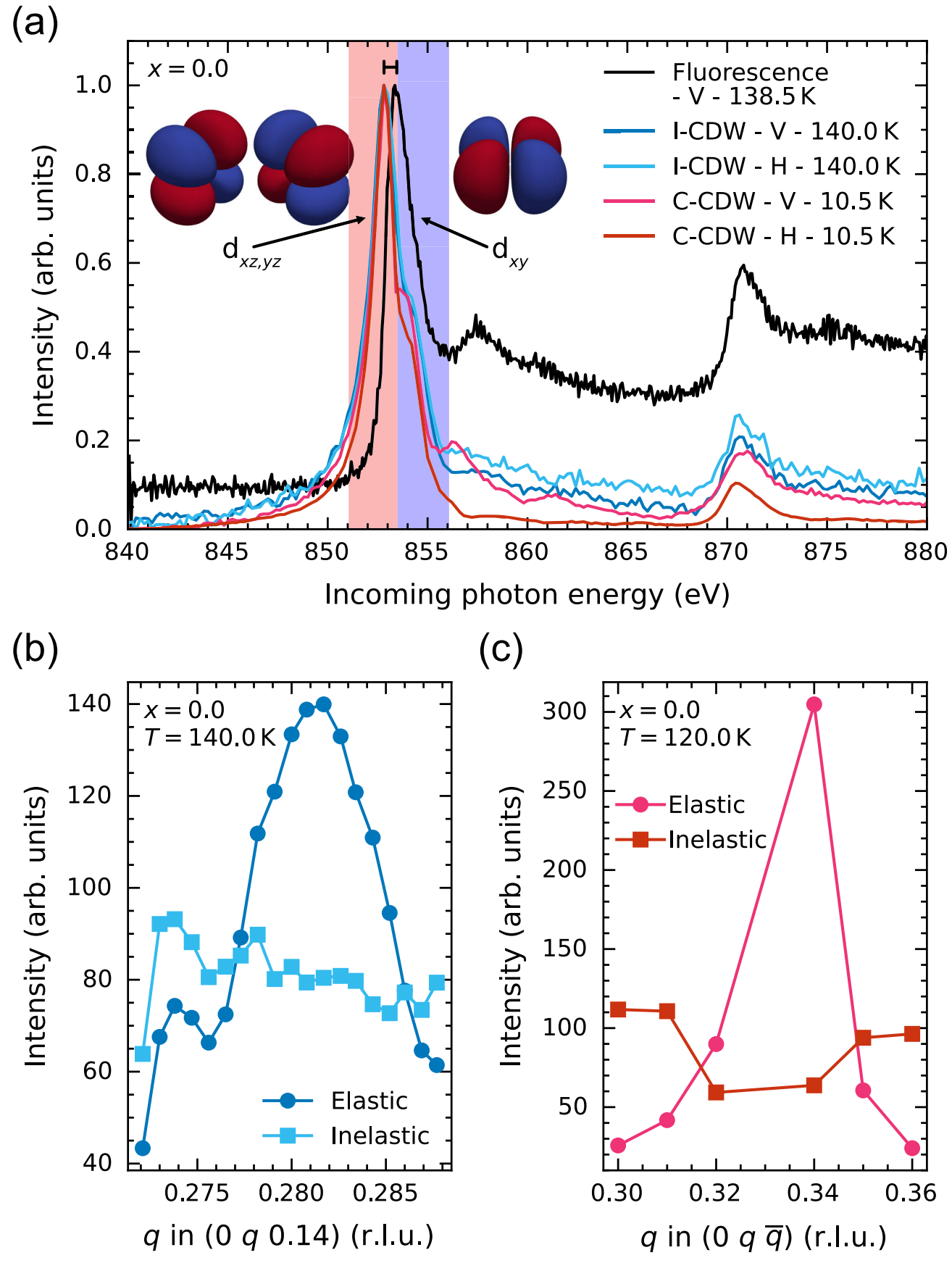

FIG. 2. (a) Incoming photon energy dependence of the scattering signal of the I-CDW and C-CDW. The data are shown for incoming vertical and horizontal polarization at an azimuthal rotation of 0°. For comparison measurements of the fluorescence, which mainly show the x-ray absorption, are included. The energy regions of the $d_{xz,yz}$ and $d_{xy}$ are highlighted in light red and light blue, respectively. (b),(c) Integrated intensity of the elastic ($-0.1$ eV $\leq E \leq 0.1$ eV) and inelastic ($E \leq -0.1$ eV) spectra for different positions around the (b) I-CDW and (c) C-CDW.

To examine whether excitations (of charge or orbital origin) contribute to the CDW signals, we performed RIXS measurements in the vicinity of the I-CDW and C-CDW peaks. The integrated elastic and inelastic intensity measured across these peaks are shown in Figs. 2(b) and 2(c), respectively. The inelastic component remains mainly flat, while the elastic intensity shows clearly the I-CDW and C-CDW peaks, confirming that the observed CDW signal is purely elastic. A full map of the RIXS signal as a function of the incoming and outgoing photon energies reveals, besides the elastic peak, only x-ray fluorescence besides the elastic peak (see Fig. S8 in SM [32]).

*Azimuthal dependence*—We next examine the azimuthal dependence of the CDW scattering signals. All measurements were performed at the CDW resonance maximum (852.8 eV) using $\theta$–$2\theta$ scans as a function of the azimuthal angle $\varphi$ that rotates the sample around the scattering vector [see Fig. 1(c)]. For each azimuth, scans were carried out with both vertically and horizontally polarized incident light. The azimuthal angle zero was defined such that the *b*-*c* plane lies in the scattering plane.

Representative $\theta$–$2\theta$ scans of the I-CDW peak are shown in Figs. 3(a) and 3(b) for vertical and horizontal polarization, respectively. At selected azimuthal angles, additional measurements with intermediate linear polarizations between horizontal (0°) and vertical (90°) were performed at the UE46-PGM-1 beamline. The integrated intensity of the I-CDW peak on Sample 1 for three selected azimuths is presented in Fig. 3(c). Complete datasets of $\theta$–$2\theta$ scans and integrated intensities for all linear polarizations in both the I-CDW and C-CDW phases are provided in SM (Figs. S5 and S6 [32]).

We begin with the I-CDW. The raw $\theta$–$2\theta$ scans [in Figs. 3(a) and 3(b)] and integrated intensities [Fig. 3(c)] reveal a clear dependence on both the azimuth and incident polarization. Additional insight is gained by considering the ratio of intensities measured with incoming vertical $I_v$ and horizontal $I_h$ polarization, which helps suppressing geometric effects such as misalignment or surface roughness. The resulting azimuthal dependence of this ratio is shown in Fig. 3(d) and exhibits a distinct fourfold modulation with nearly equal peak heights and spacing. Across the full azimuthal range, the intensity with incoming vertical polarization remains similar to or larger than that with incoming horizontal polarization. Comparison between the pristine and $x \approx 0.059$ P-substituted samples shows only minor differences in this behavior.

Before interpreting the azimuthal dependence, we introduce the expression for the measured intensity ratio. Since the outgoing polarization is not resolved in the present setup, the total intensity corresponds to the absolute square of the coherent sum of scattering amplitudes for both outgoing polarization channels and is given by

$$\frac{I_v}{I_h} = \frac{|\vec{\epsilon}^*_{s,v}\mathbf{R}\mathbf{f}^{\mathrm{res}}\mathbf{R}^T\vec{\epsilon}_{i,v}|^2 + \left|\vec{\epsilon}^*_{s,h}\mathbf{R}\mathbf{f}^{\mathrm{res}}\mathbf{R}^T\vec{\epsilon}_{i,v}\right|^2}{|\vec{\epsilon}^*_{s,v}\mathbf{R}\mathbf{f}^{\mathrm{res}}\mathbf{R}^T\vec{\epsilon}_{i,h}|^2 + \left|\vec{\epsilon}^*_{s,h}\mathbf{R}\mathbf{f}^{\mathrm{res}}\mathbf{R}^T\vec{\epsilon}_{i,h}\right|^2}, \qquad (1)$$

including the incident $\vec{\epsilon}_i$ and scattered $\vec{\epsilon}_s$ polarization vectors for vertical $\vec{\epsilon}_v$ and horizontal $\vec{\epsilon}_h$ polarization, the rotation matrix $\mathbf{R}$ accounting for the wedge needed for the C-CDW, and the resonant part of the tensor atomic form factor $\mathbf{f}^{\mathrm{res}}$ (see SM for more details [32]). The polarization vectors and the rotation matrix are determined by the scattering geometry, thus, the azimuthal dependence is governed by $\mathbf{f}^{\mathrm{res}}$. Because of the use of intensity ratios, the form factor is defined only up to an overall complex scalar, allowing one tensor component to be set to unity without loss of generality. As discussed in SM, the $\mathbf{f}^{\mathrm{res}}$ can be expanded in terms of Cartesian tensors corresponding to different multipole transitions. In the simplest case of purely dipole transitions—relevant to the present discussion—the form factor reduces to a second-rank $3 \times 3$ tensor $\mathbf{D}$. This tensor is constrained by the nature of the scattering (here charge) and by the local point group symmetry of the resonant atom (nickel), requiring $\mathbf{D}$ to be consistent with the allowed symmetry operations [38].

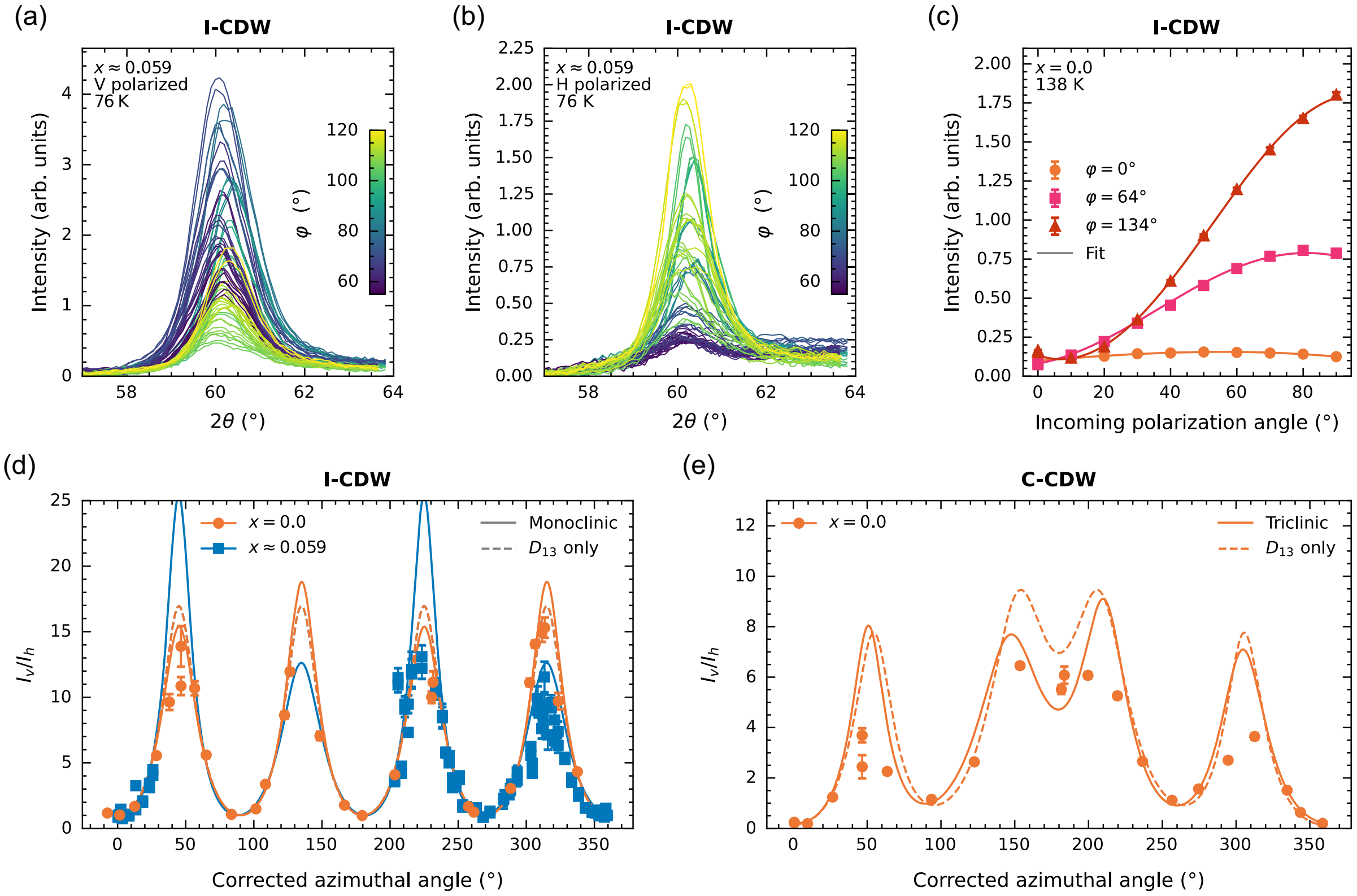

FIG. 3. (a),(b) $\theta - 2\theta$ scans for different azimuthal rotations with incoming (a) vertically polarized x-rays and (b) horizontally polarized x-rays. (c) Intensities for different incoming linearly polarized x-rays at different azimuthal angles. The measurements were performed at 138 K. (d) Azimuthal dependence of the ratio of the I-CDW scattering signal with incoming vertical polarization ($I_v$) and incoming horizontal polarization ($I_h$). The I-CDW measurements of the $x = 0.0$ sample were measured at 138 K and the measurements of the $x \approx 0.059$ sample were performed at 76 K. (e) Azimuthal dependence of the C-CDW scattering signal. The C-CDW measurements were performed at 10.5 K. For (d) and (e) two different sets of fits are shown. In the first set, all components of the dipole-dipole tensor allowed by the monoclinic and triclinic symmetry are considered for the I-CDW and C-CDW, respectively. In the second one, all terms but the dominant $D_{13}$ component of the dipole-dipole tensor were forced to zero.

The azimuthal dependence of the I-CDW scattering signal follows a clear $\sin^2(2\varphi)$ modulation, which can only be captured by a single component of the dipole-dipole tensor $\mathbf{D}$, namely $D_{13}$ (see SM [32]). Assuming $D_{13}$ to be the only nonzero component yields a good description of the observed dependence [see Fig. 3(d)]. The highest local symmetry of the Ni site compatible with a nonzero $D_{13}$, is monoclinic with the twofold axis aligned along the $y$ direction [39]—coinciding with the direction of the I-CDW wave vector. For this symmetry the dipole-dipole tensor for charge scattering takes the form

$$\mathbf{D}_{\mathrm{mono}} = \begin{pmatrix} D_{11} & 0 & D_{13} \\ 0 & D_{22} & 0 \\ D_{13} & 0 & D_{33} \end{pmatrix}, \tag{2}$$

as tabulated in Ref. [39]. In contrast, the orthorhombic symmetry proposed as the average structure in the I-CDW state [18], does not permit the $D_{13}$ component of the tensor and so cannot reproduce the observed azimuthal behavior (see SM [32]).

Given the dominance of the $D_{13}$ term, its value is fixed to 1 in the fit. See SM [32] for a detailed analysis of the individual components, which highlights the dominance of the $D_{13}$ component, as well as details of the fitting process.

The resulting fits to the azimuthal dependence are shown in Fig. 3(d). The calculated polarization dependencies,

based on the same tensor model (with fitted absolute intensity) are shown in Fig. 3(c). The fit parameters are listed in Table S-II [32].

For both samples, the monoclinic dipole-dipole tensor provides a good fit to the azimuthal and linear polarization dependence, with $D_{13}$ as the dominant component. This confirms that the local symmetry at the Ni site must be monoclinic (or lower).

Given the lack of inversion symmetry in both the proposed orthorhombic (PG: *mm*2) and low-temperature triclinic (PG: 1) structures [18], additional dipole-quadrupole contributions may be allowed. While these improve the fit for the doped sample, the large number of free parameters prevents a conclusive analysis. Lowering the symmetry to triclinic without including quadrupole terms does not significantly improve the fit. Further details are provided in SM [32].

We now turn to the azimuthal dependence of the C-CDW. The azimuthal dependence measure on pristine $BaNi_2As_2$ is shown in Fig. 3(e). Linear polarization dependencies at selected azimuths are provided in SM [32]. The data again show a four-peak structure, though the peaks differ in height and spacing compared to the I-CDW case.

Assuming $D_{13}$ as the only nonzero dipole-dipole tensor component reproduces the peak positions, indicating that the asymmetries arise primarily from the different scattering geometry [see Fig. 3(e)]. In the triclinic structure, the local Ni symmetry is 1, allowing all the components of the tensor [39]

$$\mathbf{D}_{\mathrm{tri}} = \begin{pmatrix} D_{11} & D_{12} & D_{13} \\ D_{12} & D_{22} & D_{23} \\ D_{13} & D_{23} & D_{33} \end{pmatrix}. \quad (3)$$

Fitting the data with this full tensor yields a good agreement, with $D_{13}$ again dominating, though other components contribute more significantly than in the I-CDW case. Including symmetry allowed dipole-quadrupole terms improves the fit quality (reducing the residual sum squared by nearly an order of magnitude), but the large parameter space precludes a reliable determination.

*Discussion*—The energy dependence of both the I-CDW and the C-CDW shows a clear resonant enhancement at the Ni $L_3$ edge, indicating that the observed peaks are associated with transitions between the Ni $2p_{3/2}$ and 3d orbitals. The combined energy and azimuthal dependencies provide clear and direct evidence that the resonant scattering intensity originates from an orbital order, specifically from a modulation of the Ni $d_{xz,yz}$ orbital occupation. The existence and dominance of the $D_{13}$ term demonstrate that the Ni 3d orbitals are not isotropically occupied but acquire a preferred orientation, i.e., they exhibit orbital polarization. Moreover, the shape of the resonant profile differs fundamentally from that expected for a purely structural distortion and from that of an orbital-energy modulation, as discussed in Refs. [24,27–29]. The observed resonance cannot be reproduced by a lattice-driven mechanism and is inconsistent with a mere spatial variation of orbital energies [32]. We therefore conclude that the CDW scattering arises from a genuine modulation of the Ni $d_{xz,yz}$ orbital occupation, evidencing the presence of long-range orbital order.

While the soft phonon driving the I-CDW has negligible (nonresonant) structure factor at the probed wave vector, both the I-CDW and C-CDW are also observed with hard x-ray diffraction [13], suggesting a mixed character with both electronic and lattice contributions. However, modulation of the CDW resonance due to variation in orbital occupation first requires a detailed x-ray absorption study—including substitutions with changing valence—which have not yet been reported and are beyond the scope of this Letter. Nevertheless, by comparing the resonance energies to those previously reported by NEXAFS [18], we identified the $d_{xz,yz}$ orbitals as the dominant contributors to both CDWs. The resonance aligns with the reported redistribution of spectral weight from $d_{xy}$ to the $d_{xz,yz}$, linking this orbital rearrangement to CDW formation. Notably, the onset of this orbital redistribution at temperatures higher above the static I-CDW transition indicates the presence of fluctuating electronic order, likely related to the nematic fluctuations of orbital origin inferred from by Raman experiments [17]. Furthermore, the dominant role of $d_{xz,yz}$ orbitals may explain the suppression of all CDW instabilities under hydrostatic pressure, where strong change in the Ni-As hybridization and therefore the Ni $d_{xz,yz}$ orbitals has been reported [14].

Focusing on the I-CDW phase, our results show that the Ni site experiences a lowering of its local (point) symmetry compared to the high-temperature phase. Since the transition is of second order, the crystal structure itself must transform through a group-subgroup relation of the high-temperature tetragonal space group, implying that the low-temperature phase belongs to one of its symmetry-reduced subgroups. Thermal expansion measurements exclude all space groups with fourfold axes and F-centered cells [18], while our azimuthal REXS data further rule out all space groups with a point group symmetry higher than monoclinic symmetry for the Ni position. This narrows the candidates to $I222$ (orthorhombic, no inversion center) and $C2/m$ (monoclinic, with inversion center) as the two space groups with the highest symmetry fulfilling all constraints. Both the original tetragonal and low-temperature triclinic [40] phases exhibit an inversion center, making $C2/m$ the more likely space group despite being of lower symmetry. All possible structures allow for dipole-quadrupole terms independently. Reducing the symmetry of the space group will also lead to twin formation. However, no scattering signal is expected at the investigated I-CDW position (orthorhombic), or the influence remains small because the structural distortion is small. In any case, twin formation cannot result in a situation that appears to have a lower symmetry than the actual crystal.

The lowered local and crystal symmetry strongly motivates further structural studies of $BaNi_2(As_{1-x}P_x)_2$, including the I-CDW modulation, which requires a refinement in a superspace group with at least four dimensions.

At least up to $x \approx 0.059$, we find no significant modification of the tensor atomic form factor for the I-CDW, suggesting that the local symmetry and orbital configuration remain robust across this doping range. The identification of the dominant orbitals and the symmetry constraints derived here significantly narrow the space of viable mechanisms for the proposed unconventional, orbital-driven origin of the I-CDW [15].

Comparing the I-CDW and C-CDW, we find a striking similarity in their resonant behavior. Both show nearly identical energy dependence, and their azimuthal dependencies are dominated by the $D_{13}$ component of the dipole-dipole tensor. While higher-order components are more relevant for the C-CDW, the overall orbital character of the two CDW phases appears comparable. This similarity contrasts with the substantial differences in their average crystallographic structures, distinct responses to hydrostatic pressure [14], and differing suppression under phosphorus substitution [13]. Furthermore, a soft phonon has been identified as the driving mode for the I-CDW transition [15], whereas no analogous phonon mode has been observed for the C-CDW [13], suggesting differences in their lattice dynamics despite their similar orbital signatures.

*Conclusions*—In summary, we investigated the incommensurate and commensurate charge density waves in $BaNi_2(As_{1\text{-}x}P_x)_2$ using resonant x-ray scattering. Both CDWs exhibit a pronounced resonant enhancement and a strong polarization and azimuthal dependence, providing direct evidence for orbital order. The four-peak azimuthal modulation, dominated by the $D_{13}$ component of the dipole-dipole tensor, identifies the Ni $d_{xz,yz}$ orbitals as the primary contributors to the CDW states. This orbital signature reveals that the I-CDW phase involves a lowering of the local Ni site symmetry to monoclinic or lower, with the twofold axis aligned along the CDW wave vector. The polarization-dependent REXS response confirms that the observed CDW peaks are not solely due to structural distortions but arise from a genuine modulation of the electronic charge density associated with orbital degrees of freedom. Our findings establish orbital polarization of the Ni $d_{xz,yz}$ states as the microscopic driver of charge order in $BaNi_2(As_{1\text{-}x}P_x)_2$, linking the CDWs to lattice and transport anisotropies observed in related measurements [8–18,41]. Beyond this material, they highlight orbital-lattice coupling as a key ingredient in the broader family of Ni-based superconductors, motivating complementary studies to probe this mechanism across the superconducting dome.

*Acknowledgments*—We acknowledge the European Synchrotron Radiation Facility (ESRF) for provision of synchrotron radiation facilities under proposal no. HC-5233 and we would like to thank N. Brookes for assistance and support in using beamline ID32. We thank the Helmholtz-Zentrum Berlin für Materialien und Energie for the allocation of synchrotron radiation beam time at BESSY II under the proposal no. 221-10813-ST. We acknowledge the funding by the Deutsche Forschungsgemeinschaft (DFG; German Research Foundation) Project-ID 422213477-TRR 288 (Project B03).

*Data availability*—The data that support the findings of this article are openly available [42,43].